\documentclass{eas}
\usepackage{graphicx}
\begin{document}

\title{The DRIFT Directional Dark Matter Experiments} 
\runningtitle{The DRIFT Experiments}
\author{E.~Daw} \address{Department of Physics and Astronomy, University of Sheffield, UK}
\author{A.~Dorofeev} \address{Department of Physics, Colorado State University, Fort Collins, CO 80523, USA}
\author{J.R.~Fox} \address{Department of Physics, Occidental College, Los Angeles, CA 90041, USA}
\author{J.-L.~Gauvreau} \sameaddress{3}
\author{ C.~Ghag} \address{School of Physics and Astronomy, University of Edinburgh, Edinburgh, EH9 3JZ, UK}
\author{L.J.~Harmon} \sameaddress{3}
\author{J. L.~Harton} \sameaddress{2}
\author{M.~Gold} \address{Department of Physics and Astronomy, University of New Mexico, NM 87131, USA}
\author{E.R.~Lee} \sameaddress{5}
\author{D.~Loomba} \sameaddress{5}
\author{E.H.~Miller} \sameaddress{5}
\author{A.St.J.~Murphy} \sameaddress{4} 
\author{S.M.~Paling} \sameaddress{1}
\author{J.M.~Landers} \sameaddress{3}
\author{N.~Phan} \sameaddress{5}
\author{M.~Pipe} \sameaddress{1}
\author{K.~Pushkin} \sameaddress{3}
\author{M.~Robinson} \sameaddress{1}
\author{S.W.~Sadler} \sameaddress{1}
\author{D.P.~Snowden-Ifft} \sameaddress{3}
\author{N.J.C.~Spooner} \sameaddress{1}
\author{D.~Walker} \sameaddress{1}
\author{D.~Warner} \sameaddress{2}
\begin{abstract}
The current status of the DRIFT (Directional Recoil Identification From Tracks) experiment at Boulby Mine is presented, including the latest limits on the WIMP spin-dependent cross-section from $1.5$ kg days of running with a mixture of CS$_2$ and CF$_4$. Planned upgrades to DRIFT IId are detailed, along with ongoing work towards DRIFT III, which aims to be the world's first $10$ m$^3$-scale directional Dark Matter detector.
\end{abstract}
\maketitle
\section{Introduction}
The DRIFT experiment has been operational at the Boulby Underground Laboratory in the UK since 2001, and remains the only m$^3$-scale directional Dark Matter search effort in the world (Ahlen {\em et al.\/} \cite{Ahlen2010}). The current iteration (DRIFT II-d) collected $47.4$ days of live-time data in 2009/10, using a mixture of CS$_2$ and CF$_4$ target gas to probe spin dependent (SD) WIMP-proton interactions. An unblind analysis of this data produced a limit on the WIMP-proton SD cross-section competitive with those set by direct detection experiments (Daw {\em et al.\/} \cite{Daw2010}). Concurrently, R\&D efforts are underway in both the UK and the USA to develop the next generation of DRIFT detectors.
\section{DRIFT IId}
\subsection{Technology Overview}
\begin{figure}[htb]
\begin{center}
\includegraphics[width=0.6\textwidth]{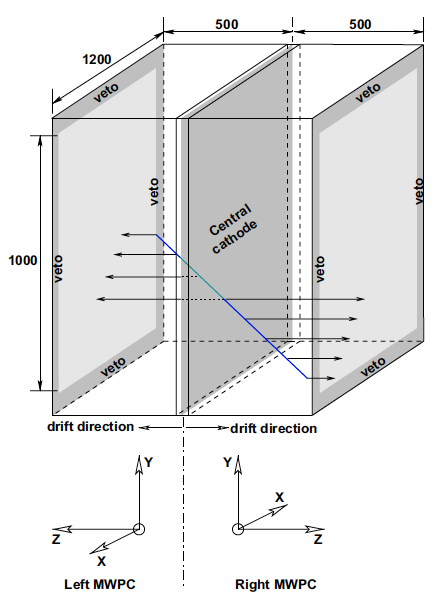}
\caption{Diagram of the DRIFT detector. Reproduced from Burgos {\em et al.\/} \cite{Burgos2008}.}
\label{fig:drift_schematic}
\end{center}
\end{figure}

The DRIFT detectors are comprised of two back-to-back negative ion time projection chambers (TPCs), each with a drift length of $50$~cm, housed inside a stainless steel vacuum vessel and encased in a neutron shield of polypropylene pellets $>67$~cm thick on all sides. The $1100$~m rock overburden ($2800$ m.w.e) of the Boulby lab ensures that the detector is not subject to cosmogenic neutrons (Robinson {\em et al.\/} \cite{Robinson2003}). The two detectors share a $-30$~kV central cathode which, coupled with the grounded anode of the readout plane, defines a  drift field of $624$~V cm$^{-1}$ in the fiducial volume. A field cage of stainless steel rings ensures a uniform electric field, which is necessary in order to preserve the spatial information of the ionization tracks created by particle interactions in the fiducial volume. Tracks are drifted toward two identical multi wire proportional counter (MWPC) readout planes either side of the central cathode, themselves comprised of a grounded anode plane of $512$ $20$ $\mu$m wires with $2$~mm pitch, sandwiched between two $2757$~V grid planes of $512$ $100$ $\mu$m wires at a distance of $1$~cm from, and orthogonal to, the anode wires, also with $2$~mm pitch (see figure \ref{fig:drift_schematic}). The vacuum vessel is filled with a 30:10 partial pressure mixture of CS$_2$:CF$_4$ at a total pressure of $40$~Torr; the CF$_4$ providing a half-integer spin target for spin-dependent interactions, whilst the moderately electronegative CS$_2$ acts as an electron transport gas. The total fluorine target mass available for SD WIMP-proton interactions was $31.8$~g. 

Particle interactions in the detector volume ionize target gas molecules, and the liberated electrons are captured by CS$_2$ molecules, forming negative ions. These ions are drifted to the MWPC where they are are stripped of their extra electrons, which avalanche in the high field region close to the anode wires, producing a voltage pulse proportional to the incident charge. Both the anode and the inside grid planes are read out in a grouped configuration, whereby every eighth wire is joined together and read out as a single channel. In this way, any $16\times16$~mm section of the x-y plane can fully contain expected nuclear recoil tracks from WIMP interactions, reducing readout costs at the expense of information about the absolute position of events (Alner {\em et al.\/} \cite{Alner2005}). The edge wires of both the anode and grid planes are grouped together into two veto regions, which are read out on two separate channels and used to reject events originating outside the fiducial volume.

Charge on the wires is amplified by a set of Cremat CR-111 charge-sensitive preamplifiers, mounted inside the vacuum vessel. The signals leave the vessel via a BNC feedthrough plate, and are shaped by a Cremat CR-200 Gaussian shaping amplifier. Finally, a high-pass filter consisting of a $2.2$~$\mu$F capacitor in parallel with a $50\Omega$ resistor removes low frequency periodic baseline fluctuations, and enables the DAQ to operate at a fixed threshold trigger.
\subsection{Background Reduction}
\begin{figure}[htb]
\begin{center}
\includegraphics[width=0.8\textwidth]{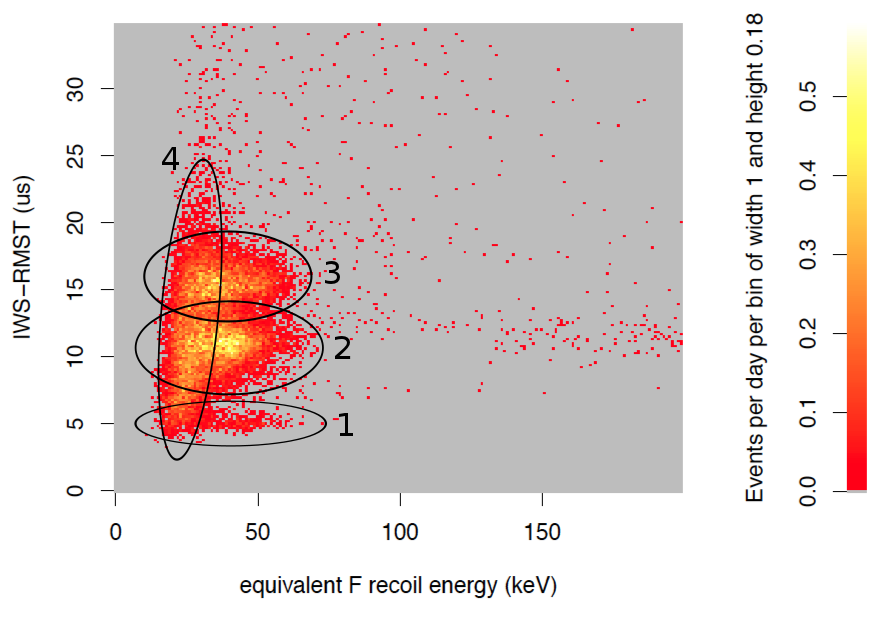}
\caption{DRIFT's four main backgrounds, revealed prior to full cuts. The events in region 1 represent sparks in the MWPC, which have characteristically short RMST and high risetimes. Region 2 is populated by recoils of radon daughter nuclei as they undergo $\alpha$ decay on the surface of wires in the MWPC. The same process accounts for the events in region 3, which this time originate on the central cathode. Region 4 contains low-energy $\beta$ and $\gamma$ events.}
\label{fig:4backgrounds}
\end{center}
\end{figure}

Four main sources of background have been identified in DRIFT, and a set of discriminants developed in order to remove them from the signal region. Figure \ref{fig:4backgrounds} shows the distribution of these background populations in RMST-Recoil Energy Space, where RMST is the voltage-weighted root mean square time of an event relative to the mean time of the event. Contributions to the RMST of an event come from both the `real' z-extent of the recoil track, and also the measured longitudinal diffusion, which increases with distance from the detector plane and dominates for the short tracks expected from WIMP-induced recoils.

Sparks (region 1, figure \ref{fig:4backgrounds}) appear as impulse events, and are therefore removed by requiring individual pulses' FWHM $> 24 \mu$s, slightly longer than the shaping time of the amplifiers, and also a risetime $> 7\mu$s. The same cuts remove a significant fraction of the events in region 4, the residual population of which defines the low-energy cutoff of the signal region. Events in region 2 occur in the MWPC and consequently have a different characteristic pulse shape from the WIMP or neutron events in which we are interested. The source of these events is probably decays of the long-lived Rn daughter, $^{210}$Pb on the surface of the MWPC wires.

DRIFT's only remaining background comes from decaying radon progeny, which have been created in nuclear decays of radon gas inside the vacuum chamber before being drifted to, and `plating out' on the cathode wires. Alpha particles are easily identified in DRIFT (Snowden-Ifft {\em et al.\/} \cite{SnowdenIfft2004}), however in the event that the alpha particle from the progeny's subsequent decay is not detected (for example, because it has been buried in the wire itself), then the detector sees only the recoiling daughter nucleus, which is very difficult to distinguish from a WIMP-induced nuclear recoil. This class of events are dubbed `untagged Radon Progeny Recoils' (RPRs); untagged, because the accompanying $\alpha$ particle in the decay is not seen and therefore not available for `tagging' the event (Burgos {\em et al.\/} \cite{Burgos2007}). Measurements of radon emanation from various materials in the DRIFT detector were performed, leading to their replacement with low-radon alternative materials. Analysis of a class of events that produce an unambiguous radon signal in the detector showed that a $90$\% reduction in rate of radon emanation was achieved by the materials substitution. Removal of $^{210}$Pb from the wires by means of nitric acid etching further reduced the rate of radon-induced events, to a value of $4\%$ of the original.

\begin{figure}[htb]
\begin{center}
\includegraphics[width=0.7\textwidth]{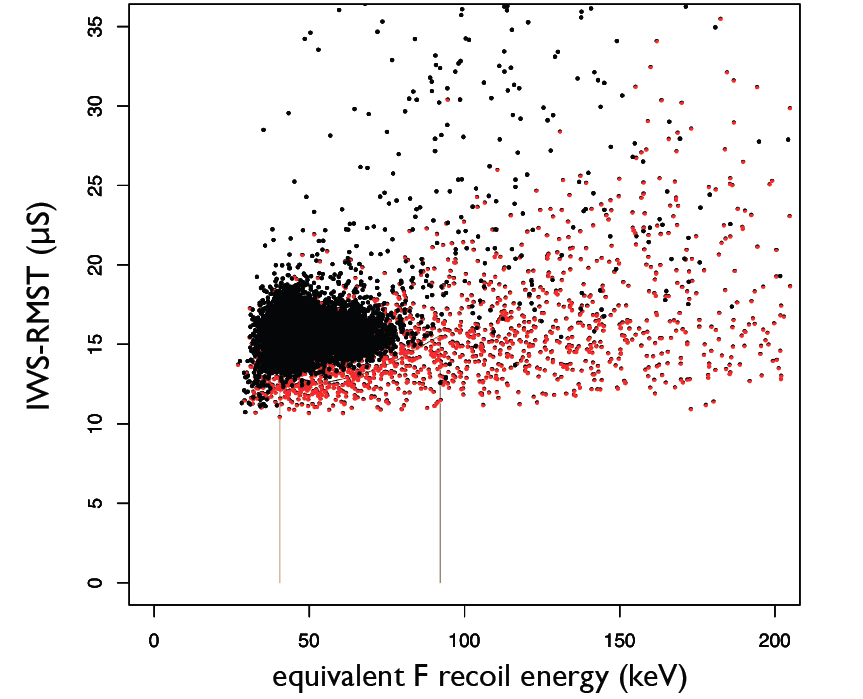}
\caption{Defining the signal region: the effect of the final residual background (black points) cut in RMST-Energy space on neutron (red points) efficiency. Events outside the tan wedge shape are cut.}
\label{fig:sig_region}
\end{center}
\end{figure}

Complementary to the physical changes to the detector, analysis cuts have been developed to remove radon-induced events in software. A high RMST means that the negative ions have drifted a long distance before reaching the MWPC, and therefore high RMST events are rejected as having originated from the central cathode (cathode RPR events). Very short pulses with a fast risetime are cut as sparks or MWPC events, however there still remains a population of low energy, moderate RMST events which can only be removed by making costly cuts into the expected WIMP parameter space (modelled by fast neutron events from a $^{252}$Cf source), as shown in figure \ref{fig:sig_region}. Further, the RMST cut is no help against events where the RPR emerges from the cathode wire with most of its velocity in the x-y plane, since the small initial z-component allows the longitudinal diffusion to mimic a z-directed event, and therefore evade the cut.

\begin{figure}[htb]
\begin{center}
\includegraphics[width=0.7\textwidth]{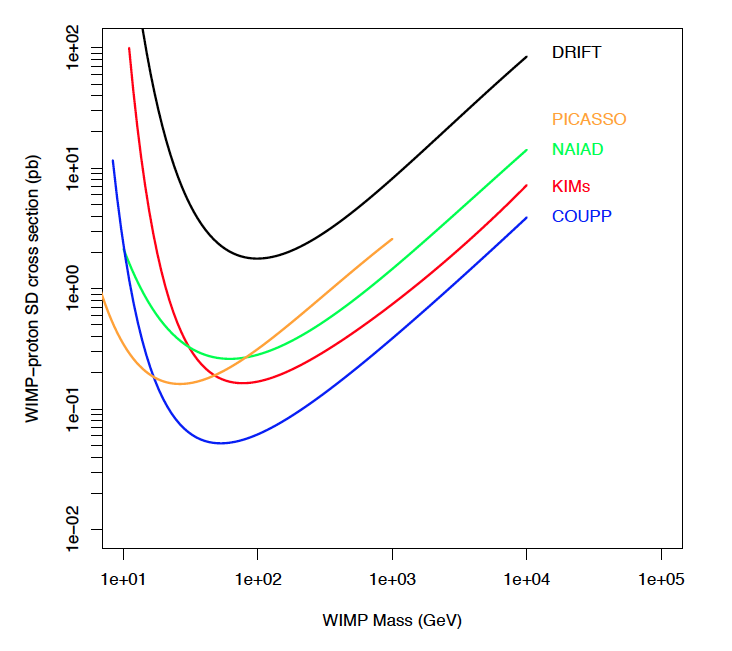}
\caption{Spin-dependent WIMP-proton limits from $1.5$~kg days of DRIFT-IId running, compared with those from several other experiments. WIMP and Picasso curves were calculated assuming standard halo parameters, whilst the other experiments assumed a range of parameters. Reproduced from Daw {\em et al.\/} \cite{Daw2010}.}
\label{fig:SDlimit}
\end{center}
\end{figure}

Figure \ref{fig:SDlimit} shows a comparison of DRIFT's recent WIMP-proton spin-dependent cross-section limit, with those from several direct detection experiments. DRIFT is the only experiment with directional sensitivity shown here. 

\section{Towards DRIFT III: ongoing R\&D with DRIFT II}
\subsection{Thin Film Cathode}
In order to mitigate the problem of untagged RPRs, a $0.9$ $\mu$m thin-film Mylar cathode plane was developed which was predicted to improve RPR discrimination by $40\times$, allowing the alpha particles from radon daughter decays to enter the fiducial volume and thereby provide a means to tag and remove these events (Miller \cite{Miller2010}). Upon installation, a substantial improvement of  $14$\% was achieved, based on the double ratio of `tagged RPR' to total background events for wire/thin-film cathodes. The discrepancy between this and the expected reduction is attributed to low energy alphas that have lost most of their energy in the film and emerged into the fiducial volume, masquerading as recoils.

The mean RMST of RPR events reduced with the introduction of the new cathode, which can be understood by considering the geometry of the thin cathode. Due to its small thickness, only RPR events which occur at very acute angles with respect to the cathode plane have a significant chance to bury their $\alpha$ particle in the cathode. 

\subsubsection{Z-fiducialisation}
The current method of z-fiducialisation, using RMST as a proxy for z-position in order to reject event occurring close to the cathode, is not ideal due to the fact that contributions to RMST come from both longitudinal diffusion (z-dependent) and a recoil's true $\Delta$z (z-independent). A new technique for z-fiducialisation based on measuring the positive ions drifting back towards the cathode has been demonstrated on the Mini-DRIFT experiment at the University of New Mexico. Measuring the difference between the time of negative ion arrival at the anode, and positive ion arrival time at the cathode yielded knowledge of an event's absolute z-position, thereby enabling true z-fiducialisation to take place.

There are significant engineering challenges associated with true z-fiducialisation. Detecting positive ions rather than electrons precludes any kind of gas amplification, therefore the signals under scrutiny are on the order of $1000$ charge units. This produces a voltage pulse after pre-amplification that is comparable in magnitude to background noise caused by microphonic vibrations in the detector, which therefore had to be suppressed by a system of elastic suspension and acoustic shielding. A major culprit for the vibrations was found to be the field cage, therefore this was replaced with a series of copper tape rings on the inner surface of a rigid I-beam cross-section Lexan support structure. Scale up to $1$m$^3$ is planned using Epoxy-laminated Kevlar as the support structure, with amplifiers mounted on the inner surface as close to the cathode as possible, in order to minimise microphonic pickup in the connecting wires.
\section{DRIFT III}
\begin{figure}[htb]
\begin{center}
\includegraphics[width=0.6\textwidth]{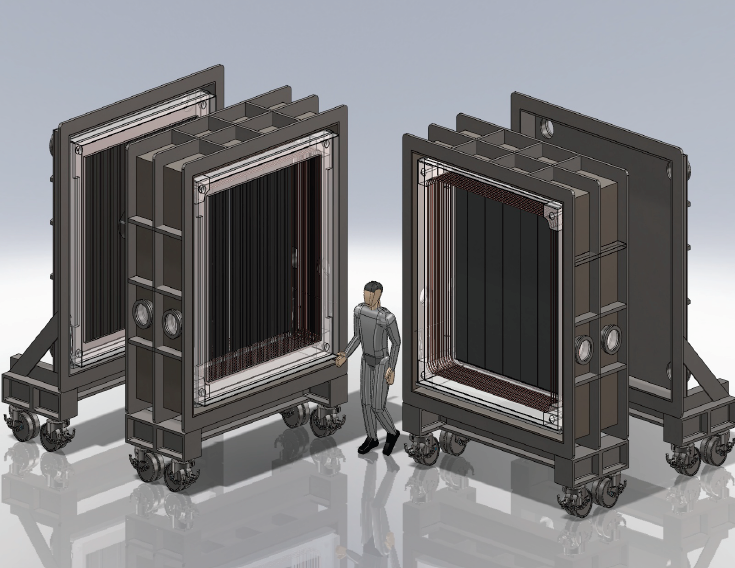}
\caption{Conceptual drawing of a proposed DTM module.}
\label{fig:dtm}
\end{center}
\end{figure}

Background rejection in DRIFT-II has now reached a point where the experiment is becoming statistics limited, highlighting the need for a scale-up. DRIFT-III is a proposed new modular directional dark matter detector, incorporating all the technological advances that have been made on previous iterations of DRIFT, and will move directional dark matter detection toward the ton scale. A cartoon of a proposed DRIFT-III Module (DTM) is shown in figure \ref{fig:dtm}. Each module consists of an instrumented $70$~kV thin-film cathode and a transparent MWPC plane capable of detecting events on either side, separated by field cage modules either side of the cathode. In this way, a large fiducial volume can be created simply by slotting in additional DTMs.
The MWPC planes will use new `resistive wire' technology, prototypes of which are currently being developed for use in DRIFT-II. Instead of measuring $\Delta$x and $\Delta$y using orthogonal anode and field wire planes, the new detectors consist of 1mm pitch alternating anode and field wires oriented in a single direction, with a recoil's extent in the direction parallel to the wires being recovered by charge-dividing readout at either end of the anode wires. 

Neutron shielding for DRIFT-III will be provided by either polypropylene pellets, as in DRIFT-II, or by high-purity water surrounding the vacuum vessels. The background neutron rate for DTM assuming $40$~g cm$^{-2}$ CH$_2$ shielding has been calculated as $< 0.4$~yr$^{-1}$ in the energy range $10-50$~keV (Carson {\em et al.\/} \cite{Carson2005}). A site for DRIFT-III at Boulby is under study. Discussions with Cleveland Potash Ltd, who run the Boulby site, suggest that the first one or two DTM modules could be housed in enlarged excavations close to the existing lab. Longer term, a further 500m of tunnels would need to be made available to accommodate the remaining 249 DTM modules necessary to reach the target mass of 1 ton. Excavations of this size can be made easily and cheapily at Boulby thanks to the relatively soft rock and ready availability of excavation machinery used for the mining process, and further discussion of future excavations are now underway.
\section*{Acknowledgements}
We acknowledge the continued support of Cleveland Potash Ltd. at Boulby Mine, as well as funding from the NSF and the STFC.

\begin{thebibliography}{}
\bibitem[2010]{Ahlen2010} Ahlen, S. {\em et al.\/} 2010, Int. J. Mod Phys A, \textbf{25}, 1
\bibitem[2005]{Alner2005} Alner, G. {\em et al.\/} 2005, Nucl. Instrum. Meth. A, \textbf{555}, 1--2
\bibitem[2007]{Burgos2007} Burgos, S. {\em et al.\/} 2007, Astropart. Phys., \textbf{28}, 1
\bibitem[2008]{Burgos2008} Burgos, S. {\em et al.\/} 2008, Nucl. Instrum. Meth. A, \textbf{584}, 1
\bibitem[2005]{Carson2005} Carson, M. {\em et al.\/} 2005, Nucl. Instrum. Meth. A, \textbf{546}, 3
\bibitem[2010]{Daw2010} Daw, E. {\em et al.\/} 2010, Arxiv preprint arXiv:1010.3027v2
\bibitem[2010]{Miller2010} Miller, E. {\em et al.\/} 2010, in \textit{Proceedings of Science - IDM 2010} 
\bibitem[2003]{Robinson2003} Robinson, M. {\em et al.\/} 2003, Nucl. Instrum. Meth. A, \textbf{511}, 3
\bibitem[2004]{SnowdenIfft2004} Snowden-Ifft, D. {\em et al.\/} 2004, Nucl. Instrum. Meth. A \textbf{516}, 1---2
\end{thebibliography}

\end{document}